\documentstyle[prd,aps,12pt,epsf,psfig]{revtex}
\begin{document}

\baselineskip=7mm
\def\ap#1#2#3{           {\it Ann. Phys. (NY) }{\bf #1} (19#2) #3}
\def\arnps#1#2#3{        {\it Ann. Rev. Nucl. Part. Sci. }{\bf #1} (19#2) #3}
\def\cnpp#1#2#3{        {\it Comm. Nucl. Part. Phys. }{\bf #1} (19#2) #3}
\def\apj#1#2#3{          {\it Astrophys. J. }{\bf #1} (19#2) #3}
\def\asr#1#2#3{          {\it Astrophys. Space Rev. }{\bf #1} (19#2) #3}
\def\ass#1#2#3{          {\it Astrophys. Space Sci. }{\bf #1} (19#2) #3}

\def\apjl#1#2#3{         {\it Astrophys. J. Lett. }{\bf #1} (19#2) #3}
\def\ass#1#2#3{          {\it Astrophys. Space Sci. }{\bf #1} (19#2) #3}
\def\jel#1#2#3{         {\it Journal Europhys. Lett. }{\bf #1} (19#2) #3}

\def\ib#1#2#3{           {\it ibid. }{\bf #1} (19#2) #3}
\def\nat#1#2#3{          {\it Nature }{\bf #1} (19#2) #3}
\def\nps#1#2#3{          {\it Nucl. Phys. B (Proc. Suppl.) } {\bf #1} (19#2) #3}
\def\np#1#2#3{           {\it Nucl. Phys. }{\bf #1} (19#2) #3}

\def\pl#1#2#3{           {\it Phys. Lett. }{\bf #1} (19#2) #3}
\def\pr#1#2#3{           {\it Phys. Rev. }{\bf #1} (19#2) #3}
\def\prep#1#2#3{         {\it Phys. Rep. }{\bf #1} (19#2) #3}
\def\prl#1#2#3{          {\it Phys. Rev. Lett. }{\bf #1} (19#2) #3}
\def\pw#1#2#3{          {\it Particle World }{\bf #1} (19#2) #3}
\def\ptp#1#2#3{          {\it Prog. Theor. Phys. }{\bf #1} (19#2) #3}
\def\jppnp#1#2#3{         {\it J. Prog. Part. Nucl. Phys. }{\bf #1} (1                                                                                                                                                                                                                                                                                                                                                                                                                                                                                                                                      {\it Nuovo Cim. }{\bf #1} (19#2) #3}
\def\r.n.c.#1#2#3{       {\it Riv. del Nuovo Cim. }{\bf #1} (19#2) #3}
\def\sjnp#1#2#3{         {\it Sov. J. Nucl. Phys. }{\bf #1} (19#2) #3}
\def\yf#1#2#3{           {\it Yad. Fiz. }{\bf #1} (19#2) #3}
\def\zetf#1#2#3{         {\it Z. Eksp. Teor. Fiz. }{\bf #1} (19#2) #3}
\def\zetfpr#1#2#3{         {\it Z. Eksp. Teor. Fiz. Pisma. Red. }{\bf #1} (19#2) #3}
\def\jetp#1#2#3{         {\it JETP }{\bf #1} (19#2) #3}
\def\mpl#1#2#3{          {\it Mod. Phys. Lett. }{\bf #1} (19#2) #3}
\def\ufn#1#2#3{          {\it Usp. Fiz. Naut. }{\bf #1} (19#2) #3}
\def\sp#1#2#3{           {\it Sov. Phys.-Usp.}{\bf #1} (19#2) #3}
\def\ppnp#1#2#3{           {\it Prog. Part. Nucl. Phys. }{\bf #1} (19#2) #3}
\def\cnpp#1#2#3{           {\it Comm. Nucl. Part. Phys. }{\bf #1} (19#2) #3}
\def\ijmp#1#2#3{           {\it Int. J. Mod. Phys. }{\bf #1} (19#2) #3}
\def\ic#1#2#3{           {\it Investigaci\'on y Ciencia }{\bf #1} (19#2) #3}
\def\tp{these proceedings}
\def\pc{private communication}
\def\ip{in preparation}
\newcommand{\TeV}{\,{\rm TeV}}
\newcommand{\GeV}{\,{\rm GeV}}
\newcommand{\MeV}{\,{\rm MeV}}
\newcommand{\keV}{\,{\rm keV}}
\newcommand{\eV}{\,{\rm eV}}
\newcommand{\Tr}{{\rm Tr}\!}
\renewcommand{\arraystretch}{1.2}
\newcommand{\be}{\begin{equation}}
\newcommand{\ee}{\end{equation}}
\newcommand{\bea}{\begin{eqnarray}}
\newcommand{\eea}{\end{eqnarray}}
\newcommand{\ba}{\begin{array}}
\newcommand{\ea}{\end{array}}
\newcommand{\bmat}{\left(\ba}
\newcommand{\emat}{\ea\right)}
\newcommand{\refs}[1]{(\ref{#1})}
\newcommand{\ler}{\stackrel{\scriptstyle <}{\scriptstyle\sim}}
\newcommand{\ger}{\stackrel{\scriptstyle >}{\scriptstyle\sim}}
\newcommand{\lag}{\langle}
\newcommand{\rag}{\rangle}
\newcommand{\ns}{\normalsize}
\newcommand{\cm}{{\cal M}}
\newcommand{\gr}{m_{3/2}}
\newcommand{\p}{\partial}
\renewcommand{\le}{\left(}
\newcommand{\ri}{\right)}
\relax
\def\321{$SU(3)\times SU(2)\times U(1)$}
\def\ord{{\cal O}}
\def\mee{m_{ee}}
\def\tl{{\tilde{l}}}
\def\tL{{\tilde{L}}}
\def\bd{{\overline{d}}}
\def\tL{{\tilde{L}}}
\def\a{\alpha}
\def\b{\beta}
\def\g{\gamma}
\def\c{\chi}
\def\d{\delta}
\def\D{\Delta}
\def\db{{\overline{\delta}}}
\def\Db{{\overline{\Delta}}}
\def\e{\epsilon}
\def\l{\lambda}
\def\n{\nu}
\def\m{\mu}
\def\nt{{\tilde{\nu}}}
\def\p{\phi}
\def\P{\Phi}
\def\sol{\Delta_{\odot}}
\def\atm{\Delta_{\mathbf{atm}}}
\def\k{\kappa}
\def\x{\xi}
\def\r{\rho}
\def\s{\sigma}
\def\t{\tau}
\def\th{\theta}
\def\om{\omega}
\def\ne{\nu_e}
\def\nm{\nu_{\mu}}
\def\snui{\tilde{\nu_i}}
\def\ehat{\hat{e}}
\def\la{{\makebox{\tiny{\bf loop}}}}
\def\ta{\tilde{a}}
\def\tb{\tilde{b}}
\def\mb{m_{1b}}
\def\mt{m_{1 \tau}}
\def\rl{{\rho}_l}
\def\meg{\m \rightarrow e \g}

\renewcommand{\Huge}{\Large}
\renewcommand{\LARGE}{\Large}
\renewcommand{\Large}{\large}
\title{
\hfill hep-ph/0202064\\ \vskip .5cm
{\Large{\bf A predictive scheme for neutrino masses}}} \vskip .5cm
\author{ Anjan S. Joshipura and  Saurabh D. Rindani\\
{\ns\it  Theoretical Physics Group, Physical Research Laboratory,}\\
{\ns\it Navarangpura, Ahmedabad, 380 009, India.}}
\maketitle
\vskip .5cm
\begin{center}
{\bf Abstract}
\end{center}

The solar and atmospheric data and possibly large value for the
effective neutrino mass in neutrinoless double beta decay
experiment together indicate that all the three neutrinos are
nearly degenerate. A verifiable texture for the neutrino mass matrix
is proposed to accommodate these results. This texture allows
almost degenerate neutrino masses two of which are exactly
degenerate at tree level. The standard model radiative corrections
lift this degeneracy and account for the solar deficit. The solar
scale is correlated with the effective neutrino mass $\mee$ probed
in neutrinoless double beta decay experiments. The model can
accommodate a large value ($\sim \ord(\eV)$) for $\mee$. Six
observables corresponding to three neutrino masses and three
mixing angles are determined in terms of only three unknown
parameters within the proposed texture.
\vskip .5cm
\noindent {\bf Introduction:} Measurement of  neutrinoless double
beta decay is of considerable theoretical importance on two
counts. Positive result would provide an unambiguous  evidence for
the non-conservation of lepton number. It would also give
direct information on neutrino masses rather than on neutrino
(mass)$^2$ differences which are probed in neutrino oscillation
experiments.

Neutrinoless double beta decay experiments measure absolute value
of an effective mass $\mee$ for the electron neutrino defined as:
\be \label{mee1}
\mee\equiv \sum U_{ei}^2 m_{\nu_i}~. \ee
$U$ denotes here the neutrino mixing matrix and $m_{\nu_i}
(i=1,2,3)$ are neutrino mass values which can take either sign.
The $\mee$ is given by the $11$ element of neutrino mass matrix in
the basis with diagonal charged lepton masses.

The experimental bound on $\mee$ is given by \cite{hm1}
$$ |\mee|\leq 0.38 ~h ~\eV~~~~~~~~{\rm at~95\%~ CL}~,$$
where $h\sim 0.6-2.8$ denotes the uncertainty in nuclear matrix
element \cite{vissani}. Recent analysis \cite{hm2} claims a
positive evidence \be \label{mee}
|\mee|\approx 0.05-0.86 \eV~~~~~~{\rm ~at~ ~95\% ~CL} \ee taking
account of uncertainty in the relevant nuclear matrix element.

The value of $\mee$ in this range if established in future
\cite{vissani} can have very important implications for particle
physics and cosmology \cite {glashow}. The implications of
neutrinoless double beta decay measurements on neutrino mass
hierarchies have been worked out in detail in a number of papers
\cite{vissani,bb,hambey}. If there are only three light neutrinos
with hierarchical masses then the result on the atmospheric
neutrino deficit ($m_{\nu_3}\leq 0.07 \eV$) and the negative
results from CHOOZ \cite{chooz} $|U_{e3}|\leq 0.12$ together imply
$\mee\leq 10^{-3}\eV$, which is substantially lower than the value
quoted in eq.(\ref{mee}). The inverted mass hierarchy
$m_{\nu_1}\approx m_{\nu 2}\gg m_{\nu_3}$ would be allowed for
$\mee$ less than or close to  the atmospheric mass scale. In
contrast, one would need almost degenerate masses \cite{deg} for
all the three neutrinos if $\mee$ is substantially higher than the
atmospheric scale.

The combined inference from the solar and atmospheric deficit and
a possible large  $\mee$ is a neutrino spectrum in which the common
mass of any two neutrinos is much larger than their (mass)$^2$
difference. One can distinguish two theoretical schemes which
allow this \cite{wolf}. Either two neutrinos have the same CP in
which case they form two almost degenerate Majorana neutrinos or
they have opposite CP and together form a pseudo-Dirac state.

A pseudo-Dirac neutrino corresponds to either of the following
structures in case of two generations, $\nu_e$ and $\nu_{x}\; (x=\mu$
or $\tau$):
\be \label{pd1}
\bmat{cc}\delta&m\\m&\delta' \emat ~,\ee
or \be \label{pd2}
\bmat{cc}a&b\\b&-a \emat ~,\ee
with $\d,\d'\ll m$ and $a\sim b$. These two textures differ from
each other both conceptually and phenomenologically. For
$\d,\d'=0$, both of them lead to neutrinos with equal and opposite
masses. The former is invariant under a global $L_e-L_x$ symmetry
which needs to be broken by non-zero $\d,\d'$ in order to generate
splitting . In contrast, the texture (\ref{pd2}) is not invariant
under any combination of $L_e$ and $L_x$. As a result, there does
not exist any symmetry to protect degeneracy and the standard
charged current interactions automatically introduce
\cite{wolf,petcov} radiative splitting among neutrinos. At the
phenomenological level, the texture in (\ref{pd1}) implies almost
maximal mixing while the mixing implied by (\ref{pd2}) is
arbitrary ($\tan 2\theta={b\over a}$). Finally, the neutrinoless
double beta decay amplitude implied by eq.(\ref{pd1}) is much
smaller than the common mass (for $\d\leq m$) while it is
comparable to the neutrino mass ($\sqrt{a^2+b^2}$) in case of
eq.(\ref{pd2}).

It follows from above that the texture in eq.(\ref{pd2}) leads to
almost degenerate pair of neutrinos with large neutrinoless double
beta decay amplitude and very small splitting. It seems ideal for
the description of neutrino masses if the latest result \cite{hm2}
are correct. The splitting introduced by the standard model
radiative correction is of $\ord({\mee m_{\tau}^2\over 16 \pi^2
M_Z^2})$. This would be in the correct range for the description
of the solar neutrino scale. One needs to generalize the basic
texture in eq.(\ref{pd2}) to three generations in order to
incorporate a solution to the atmospheric neutrino problem as
well. In the following, we suggest an economical and very
predictive scheme which provides a natural and coherent
understanding of neutrino properties required on phenomenological
ground.

\noindent {\bf Proposed Texture:} Let us consider a CP conserving
theory specified by a general $3\times 3$ real symmetric neutrino mass
matrix $M_\nu$. We require that above $M_\nu$ leads to a pseudo
Dirac neutrino. General conditions on $M_{\nu}$ under which this
happens were discussed in \cite{pdp}. In particular, the $M_{\nu}$
should satisfy

$$ tr(M_\nu)\sum_i \Delta_i=det M_\nu ~,$$ where $\Delta_i$
represents the determinant of the $2\times 2$ block of $M_{\nu}$ obtained
by blocking the $i^{th}$ row and column. While many solutions to
this constraint are possible \cite{pdp}, we study here a specific
solution which meets the requirement demanded by the observed
features of neutrino masses and mixing. The proposed solution to
the above constraint corresponds to the following neutrino mass
matrix {\it in the basis with diagonal charged leptons}:
\be \label{ansatz}M_{0\nu}= \bmat{ccc} s&t&u\\ t&-s&0\\u&0&-s\\
\emat ~.\ee
This ansatz is a direct generalization of eq.(\ref{pd2}) and
contains all the features mentioned in the $2\times 2$ case. It is given
in terms of only three parameters $s,t,u$ which after the known
radiative corrections lead to three neutrino masses and three
mixing angles making the scheme very predictive.

The $M_{0\nu }$ in eq.(\ref{ansatz}) is diagonalized to obtain
$$U_0^T~M_{0\nu}~U_0=Diag.(m,-m,-s)$$ with $$ m=\sqrt{s^2+t^2+u^2}
~.$$ The tree level mixing matrix $U_0$ is given by
\be \label{u0} U_0=\bmat{ccc}
c_1&-s_1&0\\
s_1c_2&c_2 c_1&-s_2\\
s_1s_2&s_2 c_1&c_2\\ \emat ~,\ee $s_i=\sin\theta_i$ and
$c_i=\cos\theta_i$.
The mixing angles are given by
\bea\label{theta0} \tan\theta_1&=&\sqrt{{m-s\over m+s}}~,
\nonumber \\ \tan\theta_2&=&{u\over t}~. \eea
The above mass matrix contains many desirable features at the tree
level itself.
\begin{itemize}
\item The effective neutrino mass probed in the neutrinoless
double beta decay is given by $$\mee^0=s$$
\item At the tree level, there is only one $(\rm mass)^2$ difference
which provides the atmospheric scale \be
\label{atm0}\Delta_{0A}=m^2-s^2 ~. \ee
The corresponding mixing angle ($\equiv \theta_{0A}$) coincides with
$\theta_2$ and is large when $t\sim u$:
\be \label{atang0} \sin^22 \theta_{0A}=\sin^2 2 \theta_2 ~.\ee
\item Eq.(\ref{ansatz}) already incorporates the constraint of
CHOOZ \cite{chooz} since it predicts $(U_0)_{e3}=0.$ This
prediction would receive a calculable radiative correction making
it possible to predict $U_{e3}$.
\item There is no solar splitting at this stage but the would-be solar
mixing angle is given by \be \label{solang0}
\tan^2\theta_{0S}=\tan^2\theta_{1}={\sqrt{\Delta_{0A}+(m^0_{ee})^2}-
m^0_{ee}\over \sqrt{\Delta_{0A}+(m^0_{ee})^2}+m^0_{ee}}~, \ee
It is seen that the solar mixing angle is determined in terms of
the atmospheric scale and the effective neutrino mass $m^0_{ee}$
at tree level. We will see that the radiative corrections
generate the solar splitting but do not change the above angle
appreciably.
\end{itemize}

The presence of equal and opposite neutrino masses implies a Dirac
neutrino and hence a $U(1)$ symmetry at tree level. This symmetry
is however broken by  the standard charged current interactions as
can be seen from the general expressions given in \cite{pdp}. Thus
the degenerate pair would split due to radiative interactions. This
splitting can be easily obtained \cite{pdp} using the relevant
renormalization group equations \cite{rg}. The consequences of
these equations have been discussed in a number of
papers \cite{ellislola}.

The radiatively corrected neutrino mass matrix is given by
\be \label{radmnu} M_{\nu}=I_gI_t\bmat{ccc}I_e^{\frac{1}{2}}&0&0
\\ 0&I_\m^{\frac{1}{2}}&0 \\ 0&0&I_\tau^{\frac{1}{2}}\\\emat~M_{0\nu}
~\bmat{ccc}I_e^{\frac{1}{2}}&0&0 \\ 0&I_\m^{\frac{1}{2}}&0 \\
0&0&I_\tau^{\frac{1}{2}}\\\emat ~,\ee where

$$I_\a^{\frac{1}{2}}\equiv 1+\delta_a~,$$
with \be \label{deltai}
\delta_\a\approx c\left({m_\alpha\over 4 \pi v }\right)^2 \ln{M_X\over M_Z}~.
\ee $M_X$ here corresponds to a large scale,
$c=\frac{1}{2},-\frac{1}{\cos^2\b}$ in respective cases of the
standard model and the minimal supersymmetric standard model
\cite{ellislola} and $\a=e,\mu,\tau$. $I_{g,t}$ are calculable
coefficients summarizing the effect of the gauge and the top quark
corrections.

Apart from the overall factor $I_gI_t$, the radiative corrections
are largely dominated by the $\tau$ Yukawa couplings and it is
easy to determine neutrino mixing angle and masses keeping only
$\delta_\tau$ corrections and working to the lowest order in
$\delta_\tau$. We now have
$$U^T~M_\nu~U=Diag.(m_{\nu_1},m_{\nu_2},m_{\nu_3})~,$$ with \bea
\label{masses}
m_{\nu_1}&\approx& I_g I_t ~(m+{\delta_\tau u^2\over
m+s})~,\nonumber \\ m_{\nu_2}&\approx& I_g I_t ~(-m-{\delta_\tau
u^2\over m-s})~,\nonumber \\ m_{\nu_3}&\approx& I_g I_t ~(-s-{2
\delta_\tau s t^2\over m^2-s^2})~. \eea
The tree level mixing matrix $U_0$ gets modified to a general $U$:

\be \label{u} U=\bmat{ccc}
c_1c_3&-s_1c_3&s_3\\
c_1s_2s_3+c_2s_1&c_2 c_1-s_1s_2s_3&-s_2c_3\\
-c_1c_2s_3+s_1s_2&s_1s_3c_2+s_2 c_1&c_2c_3\\ \emat ~,\ee

As before the angles, $\theta_1,\theta_2$ respectively correspond
to the solar and atmospheric mixing angles. These are now given by
\bea\label{theta}
\tan\theta_A&\approx&\tan\theta_{0A}(1-\delta_\tau), \nonumber
\\ \tan^2\theta_S&\approx&\tan^2\theta_{0S}~, \eea
where $\theta_{0A}$ (eq.(\ref{atang0}) and
$\theta_{0S}$(eq.(\ref{solang0}))denote the tree level solar
mixing angle. The effective neutrino mass $\mee^0$ is now
corrected to \be \label{meff} \mee=I_gI_t~s .\ee
The atmospheric scale also receives radiative corrections and is
now given by \be \label{atm}\Delta_{A}\equiv
\frac{1}{2}(m_{\nu_1}^2+m_{\nu_2}^2)-m_{\nu_3}^2=I_g^2I_t^2(\Delta_{0A}
+\delta_{\tau}(2 \sin^2\theta_{0A} \Delta_{0A}-\mee^{0~2} \cos 2
\theta_{0A})) ~. \ee
It is seen that all the tree level predictions receive small
radiative corrections. Hence the basic ansatz is stable against
radiaiative corrections unlike some of the ansatz considered in
\cite{ellislola}. The non-trivial effect of the radiative
corrections is generation of the solar splitting and a non-zero
$U_{e3}$: \bea \label{split}
\Delta_S&\equiv&m_{\nu_2}^2-m_{\nu_1}^2\approx
2{\delta_\tau}\mee\sqrt{\Delta_A+\mee^2}\sin ^2 2 \theta_{0A} ~,\nonumber \\
s_3&\equiv&U_{e3}\approx {\delta_\tau \mee\over
\sqrt{\Delta_A}}\sin 2 \theta_{0A}. \eea

Both $\Delta_S$ and $U_{e3}$ are correlated with the effective
neutrino mass $\mee$. This correlation is easy to understand. The
neutrino mass matrix in eq.(\ref{ansatz}) coincides with
$L_{e}-L_\m-L_\t$ symmetric structure proposed in many works
\cite{emt} when $s$ (and hence $\mee$) is zero. Since this
symmetry is also preserved by the diagonal charged lepton masses
one cannot generate the solar splitting radiatively in this case.
The presence of $s$ in the ansatz simultaneously leads to non-zero
$\mee$ and $\Delta_S$ resulting in their correlation.

\noindent{\bf Phenomenology:} It is possible to subject our ansatz
to stringent phenomenological tests since three basic parameters
predict six observables namely, $\mee$,
$U_{e3}$, ($\Delta_A,\theta_A)$ and ($\Delta_S,\theta_S)$. The
oscillation interpretation of the atmospheric neutrinos constrain
$(\Delta_A,\theta_A)$ over a narrow range
$$ \Delta_A\approx (1.5-5.0)\cdot
10^{-3}\eV^2~~~~;~~~~\sin^2 2\theta_A\approx 0.8-1 $$
The presently available results on solar neutrinos allow various
possibilities \cite{solar}. The most preferred solution based on
the global analysis of data corresponds to large mixing angle
solution (LMA) which involves relatively larger $\Delta_S$ and
$\theta_S$. The next best one corresponds to LOW and quasi-vacuum
oscillation (QVO) region. The small mixing solution is least preferred. 
The standard
model restricted fit to all the solar data rules out  this
solution at 99.73\% CL but this it is still allowed in a more
general analysis including  variations in the boron and/or hep
neutrino fluxes. We summarize below inference based on the
analysis of the observed rates in various experiments
\cite{solar}.
\vskip 1.0truecm
\begin{tabular}{|c|c|c|c|}
   \hline
   No &Solution  & $\Delta_S$&$\tan^2\theta_S$ \\
  \hline
  1 & LMA & $(1-50)\cdot 10^{-5}\eV^2$ & $0.2-0.7$ \\
  \hline
  2& SMA & $(5-10)\cdot 10^{-6}\eV^2$ & $(8-20)\cdot 10^{-4}$
  \\
  \hline
  3& LOW-QVO & $(3\cdot 10^{-9}-3\cdot 10^{-7})\eV^2$ & 0.6-2.0
  \\
  \hline
  4& VAC & $(1-2)\cdot 10^{-10}\eV^2$ & 0.2-0.6; 1.2-1.5
  \\
  \hline
\end{tabular}
\vskip .5cm
\begin{center}
{\bf Table:} The allowed ranges in parameters $\Delta_S$ and
$\tan^2\theta_S$ following from analysis of the solar rates \cite{solar}.
  All the quoted ranges are at 95\% CL. \end{center}
\vskip1.0cm Note that  the angle $\tan^2\theta_S$ is required to
be less than one for most solutions. This is crucial in
distinguishing two cases namely the standard model and the MSSM.
These two cases give different signs for $\Delta_S$ due to
difference in signs of the radiative corrections in these cases.
$\Delta_S$ is positive (negative) in case of the SM (MSSM) when
$\mee$ is positive. Conventionally, the  $\tan^2\theta_S$ is
allowed to be greater than one in the analysis of the solar data
but $\Delta_S$ is assumed positive. In the case of MSSM, positive
values of $\Delta_S$ correspond \footnote{For positive $\mee$ one
needs to reverse the role of $\nu_1$ and $\nu_2$. The relevant
$\tan^2\theta_S$ is inverse of eq.(\ref{solang0}) and is also
greater than 1.} to $\mee<0$ and negative $\mee$ implies
$\tan^2\theta_S>1$ from eq.(\ref{solang0}). Because of this reason,
the radiative corrections in MSSM are not suitable in describing
the solar data and we will restrict ourselves to the case of the
SM in the following.

We show in Fig. 1 the predicted values for $\Delta_S, \tan^2\theta_S$ and
$|U_{e3}|$ as a function of the effective mass $\mee$ for maximal
atmospheric neutrino mixing and for two extreme values for the
atmospheric mass scale. The numerical predictions include
sub-dominant corrections from the electron and muon couplings
also. We have varied $\mee$ from the maximum value of $\ord(\eV)$
to the value $\sim 10^{-3}$ which the future experiment
\cite{genius} will probe. There are two regions corresponding to
$\mee\sim 0.001-0.05 \eV$ and $\mee\sim 0.05-1.0 \eV$. The former
region corresponds to $\Delta_S\approx 10^{-9}-10^{-7} \eV^2$ and
$\tan^2\theta_S\approx 0.7-1.0$. This region encompasses the
LOW-QVO and vacuum oscillation solutions. The other region of
$\mee$ corresponds to the range reported in \cite{hm2}. In this region
only the SMA solution can be realized. As follows from
eqs.(\ref{solang0},\ref{split}), $\Delta_S$ ($\tan^2\theta_S$)
decreases (increases) with decreasing $\mee$. As a result, there
does not exist a value for $\mee$ which can simultaneously
reproduce the $\Delta_S$ and $\tan^2\theta_S$ required to realize
the most preferred LMA solution. The predicted values for $U_{e3}$
is below the expected experimental sensitivity for the entire
range in $\mee$.

\noindent {\bf Summary:} Possibly large value for the effective
majorana mass for the electron neutrino observed \cite{hm2} in
the neutrinoless double beta decay experiment points to
non-hierarchical neutrino masses which do not arise in many of the
conventional schemes such as the seesaw model or supersymmetric
model with $R$ parity violation. We have proposed an economical
ansatz for the neutrino mass matrix to describe almost degenerate
neutrino masses \cite{deg}. Justification of this ansatz from
simple symmetry may require an elaborate model. The proposed ansatz
differs in spirit from many textures discussed recently
\cite{vissani,hambey}. Unlike in these works, we discuss a
well-defined mechanism  built into our ansatz and leading to
generation of the solar scale and $U_{e3}$ through the standard
model radiative corrections. This makes the ansatz testable. The
ansatz has stringent predictions for the solar neutrino solution
which can be used to rule it out. These predictions correspond to
SMA solution if $\mee$ is in the range reported in \cite{hm2} or
to vacuum or the LOW-QVO solution if $\mee$ is much smaller.
Observation of sizable $U_{e3}> 10^{-3}$ \cite{choozm} would
also go against the ansatz.

\newpage
\begin{figure}[h]
\centerline{\psfig{figure=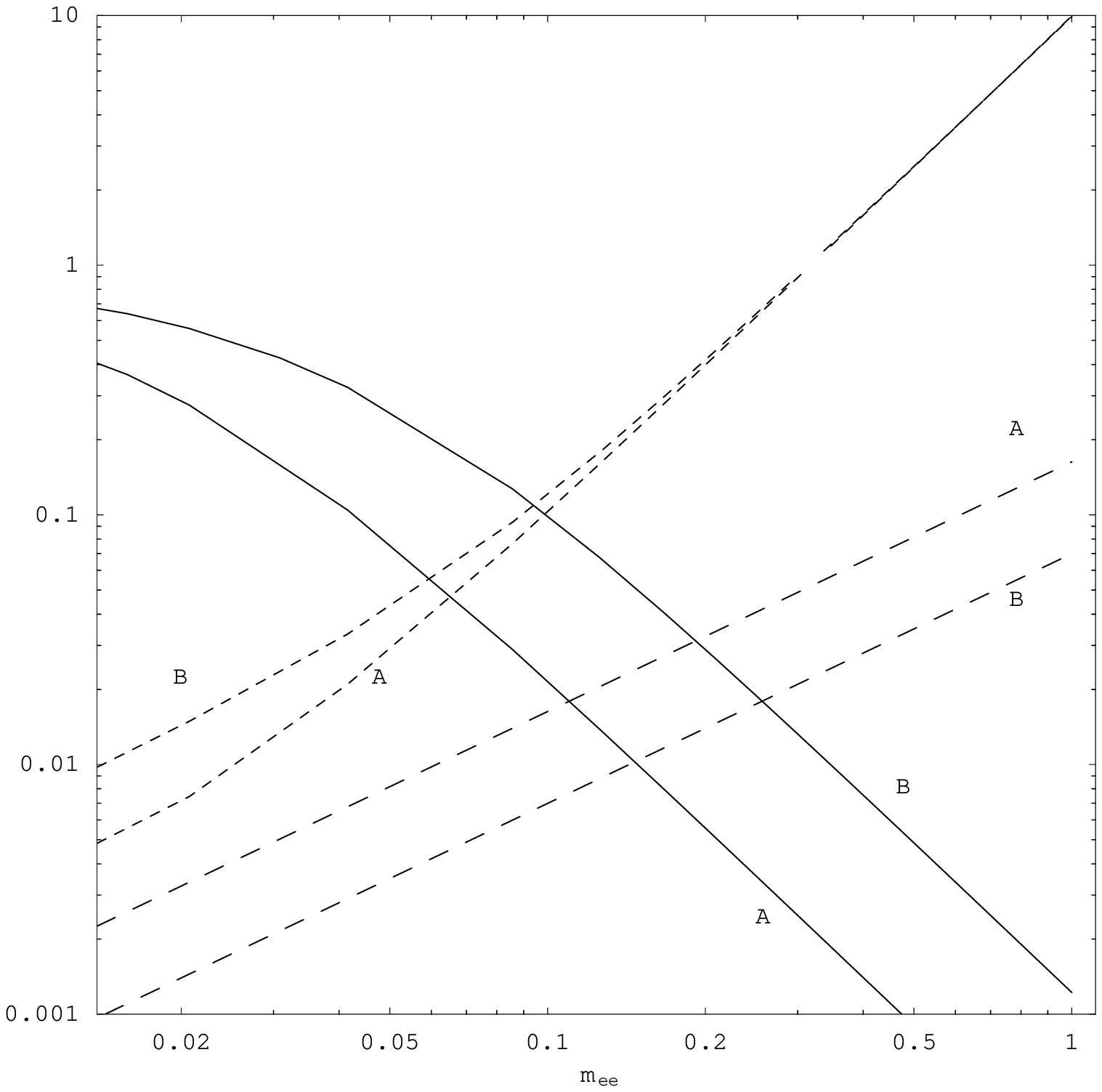,height=20cm,width=15cm}}
\vskip.25cm \caption{ $10^6 \Delta_S$ in ($\eV^2$) (dotted line),
$\tan^2\theta_S$ (solid line) and $10^3|U_{e3}|$ (dashed line) shown as a
function of $\mee$ (in $\eV$). The labels $A$ and $B$ correspond
to $\sqrt{\Delta_A}=0.03\, \eV$ and $0.07\,\eV$ respectively. The
atmospheric mixing is assumed maximal.}
\end{figure}
\end{document}